\documentclass[runningheads,fleqn]{cl2emult}
\newdimen\mathindent                              %%%%%%%%%
\mathindent = \leftmargini                        %%%%%%%%%
\usepackage{latexsym} % to have the ``\Box'' macro
\usepackage{amsfonts}

\def\lagr{{\cal L}}
\renewcommand{\a}{\alpha}
\renewcommand{\b}{\beta}

\renewcommand{\d}{\delta}
\newcommand{\pa}{\partial}
\newcommand{\g}{\gamma}
\newcommand{\G}{\Gamma}

\newcommand{\e}{\epsilon}

\newcommand{\m}{\mu}

\newcommand{\n}{\nu}

\newcommand{\s}{\sigma}

\renewcommand{\o}{\omega}

\renewcommand{\O}{{\Omega}}

\newcommand{\nn}{\nonumber}
\newcommand{\D}{{D\!\llap /\,}}
\def\be{\begin{equation}}
\def\ee{\end{equation}}
\def\bea{\begin{eqnarray}}
\def\eea{\end{eqnarray}}
\newcommand{\ft}[2]{{\textstyle\frac{#1}{#2}}}
\newcommand{\eqn}[1]{(\ref{#1})}
\begin{document}
\title*{Anti-de Sitter Supersymmetry %\newline 
}%%%%%%%%%%
\toctitle{Anti-de Sitter Supersymmetry %\protect\newline 
}
% allows explicit linebreak for the table of content
%
%
\titlerunning{Anti-de Sitter Supersymmetry}
% allows abbreviation of title, if the full title is too long
% to fit in the running head
%
\author{Bernard de Wit % \inst{1}
\and Ivan Herger  % \inst{1}}
}
\authorrunning{Bernard de Wit and Ivan Herger}
% if there are more than two authors,
% please abbreviate author list for running head
%
%
\institute{Institute for Theoretical Physics, Utrecht University\\
3508 TA Utrecht, Netherlands}
%\and Universit\'{e} de Paris-Sud,
%     B\^{a}timent 425,\\
%     F-91405 Orsay Cedex, France}

\maketitle              % typesets the title of the contribution

\begin{abstract}
We give a pedagogical introduction to certain aspects of
supersymmetric field theories in anti-de Sitter space. Among them are
the presence of masslike terms in massless wave equations, irreducible
unitary representations and the phenomenon of multiplet shortening. 
[THU-99/21; hep-th/yymmddd]
\end{abstract}

%%%%%%%%%%%%%%%%%%%%%%%%%%%%%%%%%%%%%%%%%%%%%%%%%%%%%%%%%
\section{Introduction}
Recently the study of field theory in anti-de Sitter space has
received new impetus 
by the observation that the near-horizon geometry of black branes,
which usually involves  anti-de Sitter space as a factor, is related
to a field theory associated with the 
massless modes of open strings that are attached to a certain number
$n$ of parallel Dirichlet branes, separated by small distances
\cite{Mald}. In certain   
cases there thus exists a connection between superconformal
field theories in flat space, living on the boundary of an anti-de
Sitter space-time, and gauged supergravity. The most striking example is
that of $N=4$ supersymmetric Yang-Mills theory in four space-time
dimensions with gauge group U($n$), and IIB supergravity or superstring
theory compactified on the five-dimensional sphere. 

In these lectures we intend to give a pedagogical introduction to field
theories and supersymmetry in anti-de Sitter space. The subject is
not new. Already in the thirties Dirac considered 
wave equations that are invariant under the anti-de 
Sitter group \cite{Dirac1}. Later, in 1963, he discovered the
`remarkable representation' which is now known as the singleton 
\cite{Dirac2}. Shortly afterwards there was a series of 
papers by Fronsdal and collaborators discussing the
representations of the anti-de Sitter group \cite{Fronsdal}.
Quantum field theory in anti-de Sitter space 
was studied, for instance in
\cite{AvisIshamStorey,Dusedau:1986ue}. Many new developments were
inspired by the discovery that gauged 
supergravity theories have ground states corresponding to anti-de
Sitter spacetimes
\cite{DZF,FDas,FSchwarz,dWNic,deWitNic,GatesZ,PPvN,Hull,GianPernN,GunaRomansWarner}.
This led to a study of the stability of
these ground states with respect to fluctuations of the scalar fields
\cite{BreitFreed} as well as to an extended discussion of
supermultiplets in anti-de Sitter space
\cite{BreitFreed,Heidenreich,FreedNicolai,Nicolai,GunaWarn,GunaNieuWarn}. 

In these notes we will be able to cover only a few of these topics. We
restrict ourselves to an introduction to supersymmetry in anti-de
Sitter space and discuss the presence of the so-called masslike terms
in wave equations for various fields in anti-de Sitter space. Then we will
analyze the various irreducible representations of the anti-de Sitter
isometry group, using 
a variety of techniques, and at the end we will consider the consequences for
supermultiplets. We emphasize the issue of multiplet shortening for
both multiplets of given spin and for supermultiplets. 

%%%%%%%%%%%%%%%%%%%%%%%%%%%%%%%%%%%%%%%%%%%%%%%%%%%%%%%%%%%%%%%%%%%
\section{Supersymmetry and anti-de Sitter space}\label{susy-ads}
Let us start with simple supergravity in an unspecified number of
space-time dimensions. Two important terms in any supergravity
Lagrangian are the Ein\-stein Lagrangian of general relativity
and the Rarita-Schwinger Lagrangian for the gravitino field(s), 
\be
\lagr= -\ft12e \, R(\o) - \ft12e \bar\psi_\m\,\G^{\m\n\rho}D_\n(\o)
\psi_\rho + \cdots\,,
\ee
where the covariant derivative on a spinor $\psi$ reads 
\be
D_\m(\o)\psi  = \left( \pa_\m - \ft14 \o_\m{}^{ab}\, \G_{ab}\right)\psi \,, 
\ee
and $\o_\m{}^{ab}$ is the spin-connection field defined such that
the torsion tensor (or a supercovariant version thereof) vanishes. The
action corresponding the above Lagrangian is locally supersymmetric up
to terms cubic in the gravitino field. The supersymmetry
transformations contain the terms, 
\be 
\d e_\m{}^{a} = \ft12 \bar\e \,\G^a\psi_\m\,,\qquad \d\psi_\m =
D_\m(\o) \e\,.
\ee
Extending this Lagrangian to a fully supersymmetric one is not always
possible. It may require additional fields and only when
the dimension of space-time is less than twelve does one know solutions for
interacting theories. 

Let us now include a cosmological term into the above Lagrangian as
well as a suitably chosen masslike term for the gravitino field, 
\bea
\lagr&=& -\ft12e\, R(\o) - \ft12 e \bar\psi_\m\,\G^{\m\n\rho}D_\n(\o)
\psi_\rho   \nonumber\\
&&+ \ft14 g(d-2) e \,\bar\psi_\m\G^{\m\n}\psi_\n + \ft12 g^2 (d-1)(d-2)
\,e + \cdots\,. \label{cosm-term-lagr}
\eea
As it turns out the corresponding action is still locally
supersymmetric, up to terms 
that are cubic in the gravitino field, provided that we introduce an
extra term to the transformation rules,  
\be 
\d e_\m{}^{a} = \ft12 \bar\e \,\G^a\psi_\m\,,\qquad \d\psi_\m =
\left(D_\m(\o) + \ft12 g \G_\m\right) \e\,.
\ee
This demonstrates that, a priori, supersymmetry does not forbid a cosmological
term, but it must be  of definite sign (at least, if the ground state
is to preserve supersymmetry). For a discussion see
\cite{Ferrara,deWitZwartk} and references therein. Again, to construct a fully
supersymmetric field theory is difficult and in this case there are
even stronger restrictions on the number of space-time dimensions than
in the case without a cosmological term. The Lagrangian
\eqn{cosm-term-lagr} was first written down in \cite{townsend} in four
space-time dimensions and the correct interpretation of the masslike
term was given in \cite{deser-zumino}. 

The Einstein equation corresponding to \eqn{cosm-term-lagr} reads
(suppressing the gravitino field),
\be 
R_{\m\n} -\ft12 g_{\m\n}\,R + \ft12 g^2(d-1)(d-2)\, g_{\m\n} =0\,,
\ee
which implies, 
\be 
R_{\m\n} = g^2 (d-1)\,g_{\m\n} \,,\qquad R= g^2 d(d-1)\,. 
\ee
Hence we are dealing with a $d$-dimensional  Einstein
space. The maximally symmetric 
solution of this equation is an anti-de Sitter 
space, whose Riemann curvature equals 
\be
R_{\m\n}{}^{\!ab} = 2 g^2 \, e_\m{}^{[a}\,e_\n{}^{b]}\,.
\ee
This solution leaves all the supersymmetries intact just as flat
Minkowski space does. One can verify this directly by considering the
supersymmetry variation of
the gravitino field and by requiring that it vanishes in the bosonic
background. This happens for spinors $\e(x)$ satisfying 
\be
\left(D_\m(\o) + \ft12 g \G_\m\right) \e=0\,. \label{killing-spinor}
\ee
Spinors satisfying this equation are called Killing spinors. 
Consequently also $(D_\m(\o) + \ft12 g \G_\m)(D_\n(\o) + \ft12 g
\G_\n)\e$ must vanish. Antisymmetrizing this expression in $\m$ and
$\n$ then yields the integrability condition
\be 
\Big (-\ft14 R_{\m\n}{}^{\!ab} \,\G_{ab} + \ft12 g^2 \,\G_{\m\n} \Big)
\e = 0\,,
\ee
which is precisely satisfied in anti-de Sitter space. 

Because anti-de Sitter space is maximally symmetric, it has $\ft12
d(d+1)$ isometries which constitute the
group SO$(d-1,2)$. As we have just seen, anti-de Sitter space is
consistent with  
supersymmetry. This is just as for flat Minkowski space, which has
the same number of isometries but now  
corresponding to the Poincar\'e group, and which is also
consistent with supersymmetry. The two cases are clearly related since
flat space is obtained in the limit $g\to 0$.
The algebra of the combined bosonic and fermionic symmetries
will be called the anti-de Sitter superalgebra. Note again that the
above derivation is based on an incomplete theory and in general one
will need to introduce additional fields. 
The structure of the anti-de Sitter algebra changes drastically for
dimensions $d>7$ (see \cite{nahm} and references cited therein). For
$d\leq7$ the bosonic subalgebra coincides with 
the anti-de Sitter algebra. There are $N$-extended versions, where we
introduce $N$ supersymmetry generators, each transforming as a spinor under
the anti-de Sitter group. These $N$ generators transform under a
compact group, whose generators appear in the $\{Q,\bar Q\}$
anticommutator. For $d>7$ the bosonic subalgebra can no longer be
restricted to the anti-de Sitter algebra and the algebra corresponding
to a compact group, but one needs extra bosonic generators that
transform as high-rank antisymmetric tensors under the Lorentz group. 
In contrast to
this, there exists an ($N$-extended) 
super-Poincar\'e algebra associated with flat Minkowski space of
any dimension, whose bosonic generators correspond to the
Poincar\'e group, possibly augmented with the 
generators of a compact group associated with rotations of the
supercharges.  

It is possible to describe anti-de Sitter space as a hypersurface
embedded into a $(d+1)$-dimensional embedding space. Denoting the extra
coordinate of the embedding space by $Y^-$, so that we have
coordinates $Y^A$ with $A= -,0,1,2,\ldots,d-1$, this hypersurface is
defined by 
\be
- (Y^-)^2 - (Y^0)^2 + \vec Y{}^{\,2} = \eta_{AB} \,Y^AY^B = -g^{-2}\,.
\ee
Obviously, the hypersurface is invariant under linear transformations
that leave the metric $\eta_{AB}  ={\rm diag}\,(-,-,+,+,\ldots, +)$
invariant. These transformations constitute the group SO$(d-1,2)$. The 
$\ft12d(d+1)$ generators denoted by $M_{AB}$ act on the embedding
coordinates by 
\be 
M_{AB} = Y_A{\pa\over \pa Y^B} -Y_B{\pa\over \pa Y^A}\,,
\ee
where we lower and raise indices by contracting with $\eta_{AB}$ and
its inverse $\eta^{AB}$. It is now easy to evaluate the commutation
relations for the $M_{AB}$,
\be
{[}M_{AB}, M_{CD}] = \eta_{BC}\,M_{AD} - \eta_{AC}\,M_{BD} -\eta_{BD}\,M_{AC}
+\eta_{AD}\,M_{BC}\,. \label{ads-algebra}
\ee
Anti-de Sitter space is a homogeneous space, which means that
any two points on it can be related via an isometry. It has the topology
of $S^1 \mbox{ [time] }\times {\bf R}^{d-1}$. When unwrapping 
$S^1$ one finds the universal covering space denoted by CadS, which has the
topology of ${\bf R}^{d}$. There are
many ways to coordinatize anti-de Sitter space but we will try to
avoid using specific coordinates.

On spinors, the anti-de Sitter algebra can be realized by the
following combination of gamma matrices, 
\be
M_{AB} = \ft 12 \Gamma_{AB} = \left\{ 
\begin{array}{lll}
\ft 12 \G_{ab}& \mbox{for} & A,B= a,b\,, \\[3mm]
\ft 12\G_a  &\mbox{for} & A=-\,, B= a \end{array}\right.
\ee
with $a,b= 0,1,\ldots,d-1$. Our gamma matrices satisfy the Clifford
property $\{\Gamma^a\,,\, \Gamma^b\} = 2\, \eta^{ab}\, {\bf 1}$, where
$\eta^{ab} = {\rm diag}\,(-,+,\ldots,+)$.  

The commutator of two supersymmetry transformations
yields an infinitesimal general-coordinate transformation and a tangent-space
Lorentz transformation. For example, we obtain for the vielbein,
\bea
{[}\d_1,\d_2]\,e_\m{}^a &=& \ft12 \bar\e_2\,\G^a\,\d_1\psi_\m -\ft12
\bar\e_1\,\G^a\,\d_2\psi_\m  \nn\\
&=& D_\m(\ft12\bar\e_2\,\G^a\e_1) + \ft12 g\,
(\bar\e_2\,\G^{ab}\e_1)\, e_{\m b}\,. \label{qq-comm}
\eea
Again we remind the reader of the fact that we are dealing with an
incomplete theory. For a complete theory the above result should hold
uniformly 
on all the fields (possibly modulo field equations). As before we have
ignored terms proportional to the gravitino field.  
In the anti-de Sitter background the vielbein is left invariant by the 
combination of symmetries on the right-hand side. Consequently the metric is
invariant under these coordinate transformations and we have the
so-called Killing equation, 
\be
\d g_{\m\n} = D_\m\xi_\n + D_\n\xi_\m = 0\,, \label{killing-eq}
\ee
where $\xi_\m = \ft12 \bar\e_2\,\G_\m\e_1$ is a Killing vector 
and where $\e_{1,2}$ are Killing spinors. Since $D_\m\xi_\n = \ft 12
g \bar\e_2 \G_{\m\n}\e_1$, the right-hand side of \eqn{qq-comm}
vanishes for this choice of supersymmetry parameters, and $\xi^\m$
satisfies the Killing equation \eqn{killing-eq}. As for all Killing
vectors, higher derivatives can be decomposed into the Killing vector
and its first derivative, e.g. $D_\m  (g\,\bar\e_2 \G_{\n\rho}\e_1) =
-g^2\, g_{\m[\rho}\xi_{\n]}$. The Killing vector can be decomposed
into the $\ft12 d(d+1)$ Killing vectors of the anti-de Sitter space. 

For later use we record the anti-de Sitter superalgebra, which in
addition to \eqn{ads-algebra} contains the (anti-)commutation relations, 
\bea
\{Q_\a, \bar Q_\b\} &=& - \ft12 (\G_{AB})_{\a\b} \, M^{AB} \,,\nn\\
{[}M_{AB},\bar Q_\a] &=& \ft 12 (\bar Q\, \G_{AB})_\a\,.
\label{ads-superalgebra} 
\eea
As we alluded to earlier this algebra changes its form when
considering $N$
supersymmetry generators, which rotate under the action of a compact
group. The generators of this group will then also appear on the
right-hand side of the $\{Q,\bar Q\}$ anticommutator. Beyond $d=7$
there are extra bosonic charges associated with higher-rank Lorentz
tensors. However, in
these lectures, we will mainly be dealing with the case $N=1$ and we
will always assume that $d\leq7$. 
%%%%%%%%%%%%%%%%%%%%%%%%%%%%%%%%%%%%%%%%%%%%%%%%%%%%%%%%%%%%%%%%
%%%%%%%%%%%%%%%%%%%%%%%%%%%%%%%%%%%%%%%%%%%%%%%%%%%%%%%%%%%%%%%%
\section{Anti-de Sitter supersymmetry and masslike terms}
\label{masslike}
In flat Minkowski space we know that all fields
belonging to a supermultiplet are subject to field equations with the same
mass. This must be so because the momentum operators commute 
with the supersymmetry charges, so that $P^2$ is a
Casimir operator. For
supermultiplets in anti-de Sitter space this is not longer the case,
so that masslike terms will 
not necessarily be the same for different fields belonging to the same
multiplet. This phenomenon will be illustrated below in a specific
example, namely a chiral
supermultiplet in four spacetime dimensions. Further clarification
will be given later in 
sections~\ref{Casimir} and \ref{superalgebra}.

A chiral supermultiplet in four spacetime dimensions consists of a
scalar field $A$,  
a pseudoscalar field $B$ and a Majorana spinor field $\psi$. 
In anti-de Sitter space the supersymmetry transformations of the fields are
proportional to a spinor parameter $\e(x)$, which is a
Killing spinor in the anti-de Sitter space, i.e.\ $\e(x)$ must satisfy 
the Killing
spinor equation \eqn{killing-spinor}.  We allow for two constants $a$
and $b$ in the supersymmetry transformations, which we parametrize as
follows,   
\bea
\d A &=& \ft14 \bar\e\psi\,, \qquad  \d B= \ft14 i\bar\e \g_5\psi\,,\nn\\
\d \psi &=& \D (A+ i\g_5B)\e - (a\,A + i b\,\g_5B)\e\,.
\eea
The coefficient of the first term in $\d\psi$ has been chosen
such as to ensure that $[\d_1,\d_2]$ yields the correct coordinate
transformation 
$\xi^\m D_\m$ on the fields $A$ and $B$.
To determine the constants $a$ and $b$ and the field equations of the
chiral multiplet, we consider the closure of the supersymmetry algebra
on the spinor field. After some Fierz reordering we find
\be
{[}\d_1,\d_2] \psi =  \xi^\m D_\m \psi + \ft1{16} (a-b)
\,\bar\e_2\gamma^{ab}\e_1 \, \g_{ab} \psi -\ft12  \xi^\rho \g_\rho
[\D\psi + \ft12 (a+b) \psi] \,.  \;{~}
\ee
We point out that derivatives acting on $\epsilon(x)$ occur in this
calculation at an intermediate stage and should not be suppressed in view of
\eqn{killing-spinor}. However, they produce terms proportional to $g$
which turn out to cancel in the above commutator. Now we note that the
right-hand side should constitute a coordinate transformation and 
a Lorentz transformation, possibly up to a field equation. Obviously,
the coordinate transformation coincides with \eqn{qq-comm} but
the correct Lorentz transformation is only reproduced  provided that
$a-b = 2g$. If we now denote the mass of the fermion by $m =\ft 12
(a+b)$, so that the last term is just the Dirac equation with mass
$m$, then we find
\be
a= m+g\,,\qquad b= m-g\,.
\ee
Consequently, the supersymmetry transformation of the $\psi$ equals 
\be 
\d\psi = \D (A+i\g_5 B )\e - m (A+i\g_5 B ) \e - g(A-i\g_5 B )\, 
\e\,, \label{delpsi}
\ee
and the fermionic field equation equals 
$(\D+ m)\psi =0$. The second term in \eqn{delpsi}, which is
proportional to $m$, can be accounted
for by adding an auxiliary field to the supermultiplet. The third term, which
is proportional to $g$, can be understood as a compensating
$S$-supersymmetry transformation associated with auxiliary fields in
the supergravity sector (see, e.g., \cite{deWit82}). In order to
construct the corresponding field 
equations for $A$ and $B$, we consider the variation of the fermionic
field equation. Again we have to take into account that derivatives on
the supersymmetry parameter are not equal to zero.  This yields the
following second-order differential equations,  
\bea
{}[\Box_{\rm adS} + 2g^2 - m(m-g) ]\, A&=&0\,, \nn\\
{}[\Box_{\rm adS} + 2g^2 - m(m+g) ] \,B &=&0\,,\nn\\
{}[\Box_{\rm adS} + 3 g^2 - m^2 ]\, \psi &=& 0 \,. \label{field-eqs}
\eea
The last equation follows from the Dirac equation. Namely, one
evaluates $(\D-m)(\D+m)\psi$, which gives rise to the wave operator
$\Box_{\rm adS} +\ft12 [\D,\D] - m^2$. The commutator yields the Riemann
curvature of the  anti-de Sitter space. In an anti-de Sitter space of
arbitrary dimension $d$ this equation then reads,   
\be 
{}[ \Box_{\rm adS} + \ft14 d(d-1) g^2 -m^2 ]\psi =0\,,
\ee
which, for $d=4$ agrees with the last equation of \eqn{field-eqs}. 
A striking feature of the above result is that the field equations
\eqn{field-eqs} all have different 
mass terms, in spite of the fact that they belong to the same
supermultiplet. Consequently, the role of mass is quite different in
anti-de Sitter space as compared to flat Minkowski space. This will be
elucidated later. 

For future applications we also evaluate the Proca equation for a
massive vector field,
\be
D^\mu(\pa_\m A_\n - \pa_\n A_\m) - m^2 \,A_\n = 0 \,.
\ee
This leads to $D^\m A_\m=0$, so that the field equation reads $D^2
A_\n -[D^\m,D_\n] A_\m - m^2 \,A_\n = 0$ or, in anti-de Sitter space,
\be
{}[\Box_{\rm adS} + (d-1)g^2 - m^2]\, A_\m =0 \,. 
\ee

The $g^2$ term in the field equations for the scalar fields can be
understood from the observation that the scalar D'Alembertian can be 
extended to a conformally invariant operator (see e.g. 
\cite{deWit82}),
\be
\Box + {1\over 4}{d-2\over d-1}\, R = \Box  + \ft14 d(d-2)\,
g^2 \, ,  \label{massless-scalar}
\ee
which seems the obvious candidate for a massless wave
operator for scalar fields. 
Indeed, for $d=4$, we do reproduce the $g^2$ dependence in the
first two equations \eqn{field-eqs}. Observe that 
the Dirac operator $\D$ is also conformally invariant and so is the
wave equation associated with the Maxwell field. 
%%%%%%%%%%%%%%%%%%%%%%%%%%%%%%%%%%%%%%%%%%%%%%%%%%%%%%%%%%%%%%%%%
%%%%%%%%%%%%%%%%%%%%%%%%%%%%%%%%%%%%%%%%%%%%%%%%%%%%%%%%%%%%%%%%
\section{The quadratic Casimir operator}
\label{Casimir}
To make contact between the masslike terms in the wave equations and
the properties of the irreducible representations of the anti-de
Sitter group, it is important that we establish a relation between
the D'Alembertian in anti-de Sitter space and the quadratic Casimir
operator ${\cal C}_2$ of the isometry group. 
We will use ${\cal C}_2$ later on in our discussion of the
unitary irreducible representations of the anti-de Sitter algebra. In
this section, we will  
use the $(d+1)$-dimensional flat embedding space, introduced in
section~\ref{susy-ads}, to obtain such a  relation for the scalar
D'Alembertian. In the embedding space, the latter is equal to 
to 
\be 
\Box_{d+1} = \eta^{AB} {\pa\over \pa Y^A} {\pa\over \pa Y^B}\,.
\ee
Denoting $\pa_A = \pa/\pa Y^A$ and $Y^2=\eta_{AB}Y^AY^B$, we straightforwardly
derive an expression for the quadratic Casimir operator
associated with the anti-de Sitter group SO$(d-1,2)$, 
\bea
{\cal C}_2 &=& -\ft12 \,M^{AB}\,M_{AB}   \nn\\
&=& - Y^A\,\pa^B (Y_A\pa_B -Y_B\pa_A)    \nn\\
&=& - Y^2 \,\Box_{d+1} + Y^A\pa_A\, (Y^B\pa_B + d-1)
\,. \label{embed-casimir} 
\eea
The group
SO$(d-1,2)$ has more Casimir operators but the others are of higher
order in the generators and will not play a role in the following. 
We now introduce different coordinates. We express the
$Y^A$ in terms of coordinates $X^A$, where $X^\mu=x^\m$ with $\m=
0,1,\ldots ,d-1$ and $X^-$ is defined by 
\be
X^-=\rho = \sqrt{-\eta_{AB}\,Y^AY^B}\,.
\ee
Furthermore, we require the $\rho$-dependence to be such that 
\be
Y^A(X) = \rho \,y^A(x)\,, 
\ee
so that $y^A(x)\,y^B(x)\,\eta_{AB} = -1$. With this choice of
coordinates one readily derives the following relations ($\widehat
\pa_A = \pa/\pa X^A$),
\be
\begin{array}{rclcrcl}
  \widehat \pa_- Y^A  &=& \displaystyle {1\over \rho}\,Y^A\,, &\quad&
  \widehat \pa_- Y^A \, \eta_{AB} \,Y^B &=& -\rho \,, \\[4mm]
  \widehat \pa_\mu  Y^A \,\eta_{AB} \, Y^B &=&  0 \,, & & 
  \widehat \pa_\mu Y^A \,\eta_{AB}\, \widehat\pa_-  Y^B &=&
  0\,, \\[3mm]
  \displaystyle \widehat \pa_- = {\pa\over \pa\rho}  =   
  \widehat \pa_-  Y^A\,{\pa\over\pa Y^A} &=&\displaystyle {1\over\rho} \,
Y^A\,\pa_A\,. &\quad&&&
\end{array}
\ee
In the new coordinate system the metric is given by 
\be
\widehat g_{AB} = \widehat \pa_A Y^C\,\eta_{CD}\, \widehat\pa_B  Y^D =
\pmatrix{\rho^2 g_{\m\n} & 0\cr \noalign{\vskip2mm} 0&-1\cr}\,, 
\ee
where $g_{\m\n}$ is the induced metric on the $d$-dimensional
anti-de Sitter space (with radius equal to unity). Note that $\hat g
\equiv \det \hat g_{AB} = -\rho^{2d} \det g_{\m\n}= - \rho^{2d}g$. 

The D'Alembertian of the embedding space in the new coordinates is
equal to (observe that derivatives act on all quantities on the right)
\bea
\Box_{d+1} &=& {1\over \sqrt{\widehat g}}\, \widehat\pa_{A}\, 
\widehat g^{AB}\, \sqrt{\widehat g}\; \widehat \pa_B  \nn\\
&=& {1\over \sqrt{-g}}\,{1\over\rho^d}\,\Big\{\pa_- \,\widehat
g^{--}\, \rho^d \,\sqrt{-g}\;\pa_- +
\pa_\m\, g^{\m\n} \,\rho^{d-2}\,\sqrt{- g} \;\pa_\n\Big\} \nn\\
&=& - {\pa^2 \over\pa\rho^2} - {d\over \rho} \,{\pa\over\pa\rho} +
\rho^{-2}\, \Box_{\rm adS}\,,
\eea
where $\Box_{\rm adS}$ is the D'Alembertian for the anti-de Sitter
space of unit radius. Combining this with the expression
\eqn{embed-casimir}  for the Casimir operator, we find 
\be
{\cal C}_2 = \rho^2\, \Box_{d+1}  + \rho \,{\pa\over{\pa\rho}}\Big(\rho
\,{\pa\over{\pa\rho}} + d-1\Big) = \Box_{\rm adS}\,.
\label{casimir=box}
\ee
Hence the $\pa/\pa\rho$ terms cancel as expected and the Casimir
operator is just equal to the normalized anti-de Sitter D'Alembertian
with unit anti-de Sitter radius. Note that this result cannot be used for 
other than spinless fields.

Let us now return to the wave equation for massless scalars
\eqn{massless-scalar}. According to this equation, massless $s=0$
fields lead to representations whose Casimir operator is equal to 
\be
{\cal C}_2 = - \ft 14 d(d-2)\,.
\ee
Indeed, later in these lectures we will see that the Casimir operator
for a massless $s=0$ representation in four spacetime dimensions is
equal to $-2$. 

%%%%%%%%%%%%%%%%%%%%%%%%%%%%%%%%%%%%%%%%%%%%%%%%%%%%%%%%%%%%%%%%%
\section{Unitary representations of the anti-de Sitter algebra}
\label{adSreps}
In this section we discuss unitary representations of the anti-de
Sitter algebra. For definiteness we will mainly look at the case of
four spacetime dimensions.  
We refer to \cite{Fronsdal} for some of the original work, 
and to \cite{FreedNicolai,Nicolai} where some of this
work was reviewed. In order to underline the general
features we start in $d$ spacetime dimensions. Obviously, the group
SO$(d-2,2)$ is 
noncompact. This implies that unitary representations will be infinitely
dimensional. The generators are then all anti-hermitean, 
\be
M_{AB}^\dagger = -M_{AB} \,.
\ee
Note that the covering group of SO$(d-1,2)$  has the
generators $\ft12 \Gamma_{\m\n}$ and $\ft12\Gamma_\m$. They act on spinors,
which are finite-dimensional objects. These generators, however, have
different hermiticity properties from the ones above.

The compact subgroup of the anti-de Sitter group is SO$(2)\times {\rm
SO(}d-1)$ corresponding to rotations of the compact anti-de Sitter time
and spatial rotations. It is convenient to decompose the $\ft12
d(d+1)$ generators as follows. First, the generator $M_{-0}$ is related
to the energy 
operator when the radius of the anti-de Sitter space is taken to
infinity. The eigenvalues of this generator, which is associated with
motions along the circle, are quantized in integer units in order to
have single-valued functions, unless one goes to the covering space
CadS. So we define the energy operator $H$ by 
\be 
H = -i M_{-0} \,.
\ee
Obviously the generators of the spatial rotations are the operators
$M_{ab}$ with $a,b= 1,\ldots ,d-1$. Note that we have changed
notation: here and henceforth the indices $a,b,\ldots$ refer only to
spacelike indices.  The remaining $2(d-1)$ 
generators $M_{-a}$ and $M_{0a}$ are combined into pairs of mutually
conjugate operators, 
\be 
M_a^\pm= -i M_{0a} \pm  M_{-a}\,,
\ee
and we have $(M_a^+)^\dagger = M_a^-$. The anti-de Sitter commutation
relations now read
\bea \label{ads-dec-algebra}
{[}H, M_a^\pm] &=& \pm M^\pm_a\,,\nonumber \\[1mm]
{[}M_a^\pm,M_b^\pm ]&=& 0\,,\nn\\[1mm]
{[} M_a^+ , M_b^-] &=& -2 (H\, \d_{ab} + M_{ab})\,.
\eea
Obviously, the $M_a^\pm$ play the role of raising and lowering operators:
when applied to an eigenstate of $H$ with eigenvalue $E$, application
of $M^\pm_a$ yields a state with eigenvalue $E\pm 1$.

In this section we restrict ourselves to the bosonic case. Nevertheless, let us
already briefly indicate how some of the other (anti-)commutators
of the anti-de Sitter superalgebra decompose
c.f. \eqn{ads-superalgebra},  
\bea \label{ads-dec-superalgebra}
\{Q_\a\,, Q^\dagger_\b\} &=& H\,\d_{\a\b} - 
\ft12 iM_{ab}\,(\G^a\G^b\G^0)_{\a\b} 
 \nonumber \\[1mm]
&& + \ft12(M^+_a\,\G^a\,(1+i\G^0) + M^-_a\,\G^a \,(1-i\G^0))_{\a\b}
\,,\nonumber\\[1mm]
{[}H\,, Q_\a ] &=& -\ft12 i (\G^0\,Q)_\a\,, \nonumber \\[1mm]
{[}M^\pm_a\,, Q_\a ] &=& \mp \ft12  (\G_a(1\mp i\G^0)\,Q)_\a  \,. 
\eea
For the anti-de Sitter superalgebra, all the bosonic operators can be
expressed as bilinears of the supercharges, so that in principle one
could restrict oneself to fermionic operators only and employ the
projections $(1\pm i\G^0)Q$ as the basic lowering and raising
operators. However, this is not quite what we will be doing later in
section~\ref{superalgebra}.

%%%%%%%%%%%%%%%%%%%%%%%%%%%%%%%%%%%%%%%%%%%%%%%%%%%%%%%%%
%%%%%%%%%%%%%%%%%%% spin 0 %%%%%%%%%%%%%%%%%%%%%%%%%%%%%%
%%%%%%%%%%%%%%%%%%%%%%%%%%%%%%%%%%%%%%%%%%%%%%%%%%%%%%%%%
\setlength{\unitlength}{0.5mm}
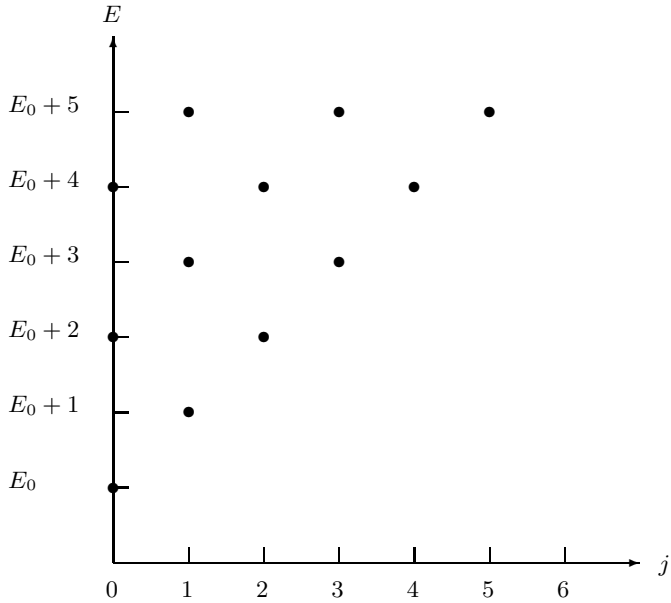
\begin{figure}[t]
\begin{picture}(260,160)(0,0)
\put(30,10){\vector(1,0){140}}
\put(175,8){$j$}
\put(30,10){\vector(0,1){140}}
\put(27,154){$E$}
\put(28,1){$0$}
\put(48,1){$1$}
\put(68,1){$2$}
\put(88,1){$3$}
\put(108,1){$4$}
\put(128,1){$5$}
\put(148,1){$6$}
\put(50,10){\line(0,1){4}}
\put(70,10){\line(0,1){4}}
\put(90,10){\line(0,1){4}}
\put(110,10){\line(0,1){4}}
\put(130,10){\line(0,1){4}}
\put(150,10){\line(0,1){4}}
\put(30,30){\line(1,0){4}}
\put(30,50){\line(1,0){4}}
\put(30,70){\line(1,0){4}}
\put(30,90){\line(1,0){4}}
\put(30,110){\line(1,0){4}}
\put(30,130){\line(1,0){4}}
\put(2,30){$E_0$}
\put(30,30){\circle*{3}}
\put(2,50){$E_0+1$}
\put(50,50){\circle*{3}}
\put(2,70){$E_0+2$}
\put(30,70){\circle*{3}}
\put(70,70){\circle*{3}}
\put(2,90){$E_0+3$}
\put(50,90){\circle*{3}}
\put(90,90){\circle*{3}}
\put(2,110){$E_0+4$}
\put(30,110){\circle*{3}}
\put(70,110){\circle*{3}}
\put(110,110){\circle*{3}}
\put(2,130){$E_0+5$}
\put(50,130){\circle*{3}}
\put(90,130){\circle*{3}}
\put(130,130){\circle*{3}}
%
%%%%%%%
%%%%%%%%%%%%%%%%%%%%%%%%%%%%%%%%%%%%%%%
%
\end{picture}
\caption{States of the $s=0$ representation in terms of the energy
  eigenvalues $E$ and the angular momentum $j$. Each point has a
$(2j+1)$-fold degeneracy. 
} 
%\vspace{-4mm}
\end{figure}
%%%%%%%%%%%%%%%%%%%%%%%%%%%%%%%%%%%%%%%%%%%%%%%%%%%%%%%%%%%%%

Let us now assume that the spectrum of $H$ is bounded from below,
\be
H\geq E_0\,,
\ee
so that in mathematical terms we are considering lowest-weight 
irreducible unitary representations. 
The lowest eigenvalue $E_0$ is realized on states that we denote by $\vert
E_0,s\rangle$, where $E_0$ is the eigenvalue of $H$ and $s$ indicates
the value of the total angular momentum operator. Of course there are
more quantum numbers, e.g.\ associated with the angular momentum
operator directed along some axis (in $d=4$ there 
are thus $2s+1$ degenerate states), but this is not important for the
construction and these quantum numbers are suppressed. Since states
with $E<E_0$ should not appear, 
ground states are characterized by the condition, 
\be
M^-_a\,\vert E_0,s\rangle = 0\,. 
\ee
The representation can now be constructed by acting with the raising
operators on the vacuum state $\vert E_0,s\rangle$. To be precise,
all states of energy $E=E_0+ n$ are constructed by an $n$-fold product
of creation operators $M^+_a$ In this way one
obtains states of higher eigenvalues $E$ with higher spin. The
simplest case is the one where the vacuum has no spin ($s=0$). For
given eigenvalue $E$, the highest spin state is given by the traceless
symmetric product of $E-E_0$ operators $M^+_a$ on the ground
state. These states are shown  in Fig.~1. 

%%%%%%%%%%%%%%%%%%%%%%%%%%%%%%%%%%%%%%%%%%%%%%%%%%%%%%%%
%%%%%%%%%%%%%%%%%%%%%% spin 1/2 %%%%%%%%%%%%%%%%%%%%%%%%
%%%%%%%%%%%%%%%%%%%%%%%%%%%%%%%%%%%%%%%%%%%%%%%%%%%%%%%%
\setlength{\unitlength}{0.5mm}
\begin{figure}[t]
\begin{picture}(260,160)(0,0)
\put(30,10){\vector(1,0){140}}
\put(175,8){$j$}
\put(30,10){\line(0,1){19}}
\put(30,31){\line(0,1){38}}
\put(30,71){\line(0,1){38}}
\put(30,111){\vector(0,1){40}}
\put(27,154){$E$}
\put(28,1){$0$}
\put(48,1){$1$}
\put(68,1){$2$}
\put(88,1){$3$}
\put(108,1){$4$}
\put(128,1){$5$}
\put(148,1){$6$}
\put(50,10){\line(0,1){4}}
\put(70,10){\line(0,1){4}}
\put(90,10){\line(0,1){4}}
\put(110,10){\line(0,1){4}}
\put(130,10){\line(0,1){4}}
\put(150,10){\line(0,1){4}}
\put(31,30){\line(1,0){3}}
\put(30,50){\line(1,0){4}}
\put(31,70){\line(1,0){3}}
\put(30,90){\line(1,0){4}}
\put(31,110){\line(1,0){3}}
\put(30,130){\line(1,0){4}}
\put(2,30){$E_0$}
\put(30,30){\circle{2}}
\put(40,30){\circle*{3}}
\put(2,50){$E_0+1$}
\put(50,50){\circle{2}}
\put(40,50){\circle*{3}}
\put(60,50){\circle*{3}}
\put(2,70){$E_0+2$}
\put(30,70){\circle{2}}
\put(40,70){\circle*{3}}
\put(70,70){\circle{2}}
\put(60,70){\circle*{3}}
\put(80,70){\circle*{3}}
\put(2,90){$E_0+3$}
\put(50,90){\circle{2}}
\put(40,90){\circle*{3}}
\put(60,90){\circle*{3}}
\put(90,90){\circle{2}}
\put(80,90){\circle*{3}}
\put(100,90){\circle*{3}}
\put(2,110){$E_0+4$}
\put(30,110){\circle{2}}
\put(40,110){\circle*{3}}
\put(70,110){\circle{2}}
\put(60,110){\circle*{3}}
\put(80,110){\circle*{3}}
\put(110,110){\circle{2}}
\put(100,110){\circle*{3}}
\put(120,110){\circle*{3}}
\put(2,130){$E_0+5$}
\put(50,130){\circle{2}}
\put(40,130){\circle*{3}}
\put(60,130){\circle*{3}}
\put(90,130){\circle{2}}
\put(80,130){\circle*{3}}
\put(100,130){\circle*{3}}
\put(130,130){\circle{2}}
\put(120,130){\circle*{3}}
\put(140,130){\circle*{3}}
%
%%%%%%%%%%%%%%%%%%%%%%%%%%%%%%%%%%%%%%%%%%%%%%
%%%%%%%%%%%%%%%%%%%%%%%%%%%%%%%%%%%%%%%%%%%%%%
%
\end{picture}
\caption{States of the $s=\ft12$ representation in terms of the energy
  eigenvalues $E$ and the angular momentum $j$. Each point has a
$(2j+1)$-fold degeneracy. The small circles denote the original $s=0$
multiplet from which the spin-$\ft12$ multiplet 
has been constructed by taking a direct product. } 
%\vspace{-4mm}
\end{figure}
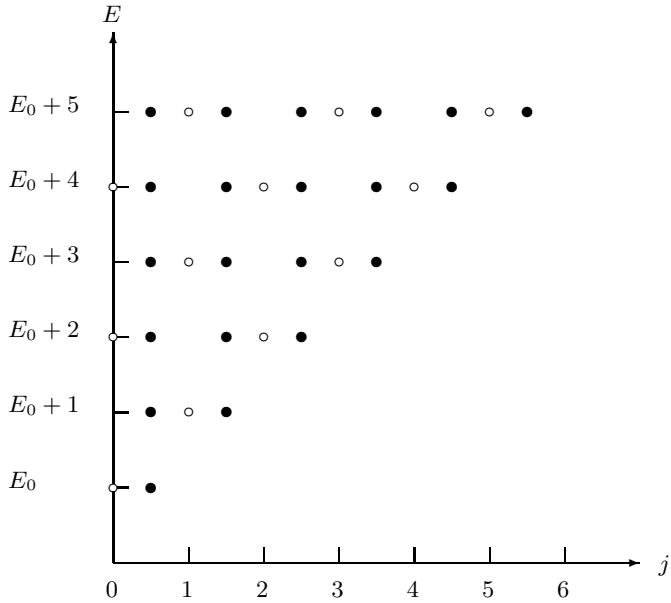
%%%%%%%%%%%%%%%%%%%%%%%%%%%%%%%%%%%%%%%%%%%%%%%%%%%%%%%%%%%%%%

Henceforth we specialize to the case $d=4$ in order to keep the aspects
related to spin simple. 
To obtain spin-$\ft12$ is trivial; one simply takes the direct product
with a spin-$\ft12$ state. That implies that every point with spin $j$
in Fig.~1 generates two points with spin $j\pm\ft12$, with the
exception of points associated with $j=0$, which will simply move to
$j=\ft12$. The result of this is shown in Fig.~2.

Likewise one can take the direct product with a
spin-1 state, but now the situation is more complicated as the
resulting multiplet is not always irreducible. In principle, each point
with spin $j$ now generates three points, associated with $j$
and $j\pm 1$, again with the exception of the $j=0$ points, which simply
move to $j=1$. The result of this procedure is shown in Fig.~3.

Let us now turn to the quadratic Casimir operator, which for $d$ spacetime
dimensions can be written as 
\bea
{\cal C}_2&=& -\ft 12 M^{AB}M_{AB}\nn\\
&=& H^2 -\ft 12 \{M_a^+,M_a^-\} - \ft12 (M_{ab})^2\nn\\
&=& H(H-d+1) - \ft12 (M_{ab})^2 -M_a^+ M_a^-\,.
\eea
Applying the last expression on the ground state 
$\vert E_0,s\rangle$ and assuming $d=4$ we derive
\be
{\cal C}_2 = E_0(E_0-3) + s(s+1) \,,
\label{4dcasimir}
\ee
and, since ${\cal C}_2$ is a Casimir operator, 
this result holds for any state belonging to the corresponding
irreducible representation. Note, that the angular momentum operator is given
by $\vec J^2 = -\ft12(M_{ab})^2$.

%%%%%%%%%%%%%%%%%%%%%%%%%%%%%%%%%%%%%%%%%%%%%%%%%%%%%%%%%
%%%%%%%%%%%%%%%%%%% spin 1 %%%%%%%%%%%%%%%%%%%%%%%%%%%%%%
%%%%%%%%%%%%%%%%%%%%%%%%%%%%%%%%%%%%%%%%%%%%%%%%%%%%%%%%%
\setlength{\unitlength}{0.5mm}
\begin{figure}[t]
\begin{picture}(260,160)(0,0)
\put(30,10){\vector(1,0){140}}
\put(175,8){$j$}
\put(30,10){\vector(0,1){140}}
\put(27,154){$E$}
\put(28,1){$0$}
\put(48,1){$1$}
\put(68,1){$2$}
\put(88,1){$3$}
\put(108,1){$4$}
\put(128,1){$5$}
\put(148,1){$6$}
\put(50,10){\line(0,1){4}}
\put(70,10){\line(0,1){4}}
\put(90,10){\line(0,1){4}}
\put(110,10){\line(0,1){4}}
\put(130,10){\line(0,1){4}}
\put(150,10){\line(0,1){4}}
\put(30,30){\line(1,0){4}}
\put(30,50){\line(1,0){4}}
\put(30,70){\line(1,0){4}}
\put(30,90){\line(1,0){4}}
\put(30,110){\line(1,0){4}}
\put(30,130){\line(1,0){4}}
\put(2,30){$E_0$}
\put(50,30){\circle*{3}}
\put(2,50){$E_0+1$}
\put(50,50){\circle*{3}}
\put(30,50){\circle*{3}}
\put(70,50){\circle*{3}}
\put(2,70){$E_0+2$}
\put(50,70){\circle*{3}}
\put(50,70){\circle{5}}
\put(70,70){\circle*{3}}
\put(90,70){\circle*{3}}
\put(2,90){$E_0+3$}
\put(30,90){\circle*{3}}
\put(50,90){\circle*{3}}
\put(70,90){\circle*{3}}
\put(70,90){\circle{5}}
\put(90,90){\circle*{3}}
\put(110,90){\circle*{3}}
\put(2,110){$E_0+4$}
\put(50,110){\circle*{3}}
\put(50,110){\circle{5}}
\put(70,110){\circle*{3}}
\put(90,110){\circle*{3}}
\put(90,110){\circle{5}}
\put(110,110){\circle*{3}}
\put(130,110){\circle*{3}}
\put(2,130){$E_0+5$}
\put(30,130){\circle*{3}}
\put(50,130){\circle*{3}}
\put(70,130){\circle*{3}}
\put(70,130){\circle{5}}
\put(90,130){\circle*{3}}
\put(110,130){\circle*{3}}
\put(110,130){\circle{5}}
\put(130,130){\circle*{3}}
\put(150,130){\circle*{3}}
%
%%%%%%%
%%%%%%%%%%%%%%%%%%%%%%%%%%%%%%%%%%%%%%%
%
\end{picture}
\caption{States of the $s=1$ representation in terms of the energy
  eigenvalues $E$ and the angular momentum $j$. Observe that there are
now points with double occupancy, indicated by the circle superimposed
on the dots. These points could combine into an $s=0$ multiplet with
ground state $\vert E_0+1,s=0\rangle$. This $s=0$ multiplet becomes
reducible and can be dropped when $E_0=2$, as is explained
in the text. The remaining points then constitute a massless $s=1$
multiplet, shown in Fig.~4. 
} 
%\vspace{-4mm}
\end{figure}
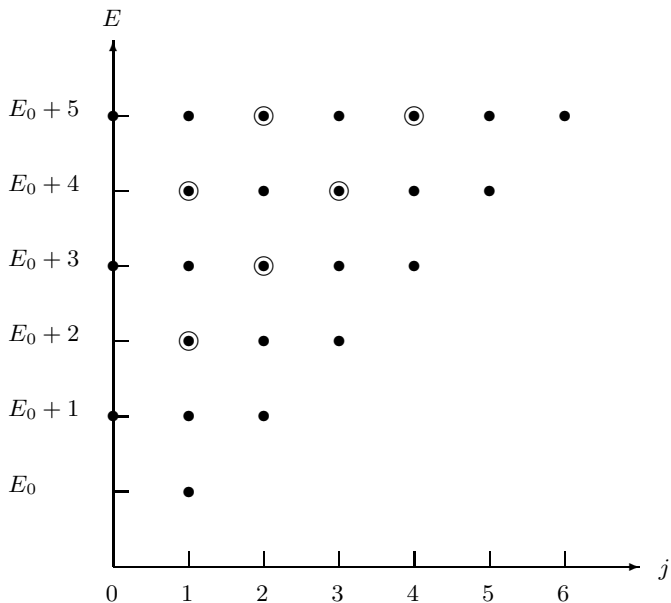
%%%%%%%%%%%%%%%%%%%%%%%%%%%%%%%%%%%%%%%%%%%%%%%%%%%%%%%%%%%%%

We can apply this result to an excited state
(which is generically present in the spectrum) 
with $E=E_0+1$ and $j=s-1$. Here, we assume that the ground state 
has $s\geq1$. In that case we find 
\bea
{\cal C}_2 &=& (E_0+1)(E_0-2) + s(s-1) - \Big\vert M_a^-  \vert
E_0+1,s-1\rangle \Big\vert^2 \nn\\
&=& E_0(E_0-3) + s(s+1)\,,
\eea
so that 
\be
E_0-s-1 = \ft12 \Big\vert M_a^-  \vert
E_0+1,s-1\rangle \Big\vert^2 \,.
\ee
This shows that $E_0 \geq s+1$ in order to have a unitary
multiplet. When $E_0= s+1$, however,  the state $\vert
E_0+1,s-1\rangle$ is itself a ground state, which decouples from the
original multiplet, 
together with its corresponding excited states. This can be
interpreted as the result of a gauge symmetry and therefore we call
these multiplets {\it massless}. Hence massless multiplets with
$s\geq1$ are characterized by 
\be 
E_0= s+1\,, \qquad \mbox{ for} \quad s\geq 1\,.
\ee
For these particular values the quadratic Casimir operator is  
\be
{\cal C}_2 = 2(s^2 -1) \,.
\label{masslesscasimir}
\ee
Although this result is only derived for $s\geq 1$, it also applies to
massless $s=0$ 
and $s=\ft12$ representations, as we shall see later. Massless $s=0$
multiplets have either $E_0=1$ or $E_0=2$, while massless $s=\ft12$
multiplets have $E_0=\ft32$. 

%%%%%%%%%%%%%%%%%%%%%%%%%%%%%%%%%%%%%%%%%%%%%%%%%%%%%%%%%
%%%%%%%%%%%%%%%%%%% massless spin 1 %%%%%%%%%%%%%%%%%%%%%
%%%%%%%%%%%%%%%%%%%%%%%%%%%%%%%%%%%%%%%%%%%%%%%%%%%%%%%%%
\setlength{\unitlength}{0.5mm}
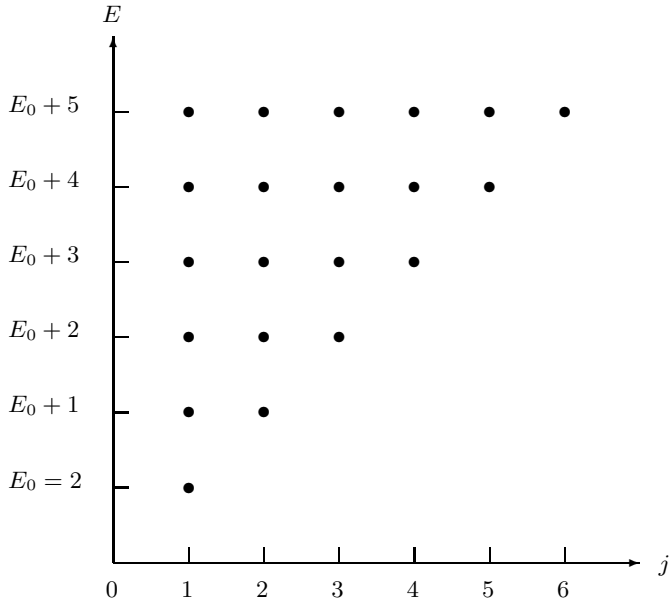
\begin{figure}[t]
\begin{picture}(260,160)(0,0)
\put(30,10){\vector(1,0){140}}
\put(175,8){$j$}
\put(30,10){\vector(0,1){140}}
\put(27,154){$E$}
\put(28,1){$0$}
\put(48,1){$1$}
\put(68,1){$2$}
\put(88,1){$3$}
\put(108,1){$4$}
\put(128,1){$5$}
\put(148,1){$6$}
\put(50,10){\line(0,1){4}}
\put(70,10){\line(0,1){4}}
\put(90,10){\line(0,1){4}}
\put(110,10){\line(0,1){4}}
\put(130,10){\line(0,1){4}}
\put(150,10){\line(0,1){4}}
\put(30,30){\line(1,0){4}}
\put(30,50){\line(1,0){4}}
\put(30,70){\line(1,0){4}}
\put(30,90){\line(1,0){4}}
\put(30,110){\line(1,0){4}}
\put(30,130){\line(1,0){4}}
\put(2,30){$E_0=2$}
\put(50,30){\circle*{3}}
\put(2,50){$E_0+1$}
\put(50,50){\circle*{3}}
\put(70,50){\circle*{3}}
\put(2,70){$E_0+2$}
\put(50,70){\circle*{3}}
\put(70,70){\circle*{3}}
\put(90,70){\circle*{3}}
\put(2,90){$E_0+3$}
\put(50,90){\circle*{3}}
\put(70,90){\circle*{3}}
\put(90,90){\circle*{3}}
\put(110,90){\circle*{3}}
\put(2,110){$E_0+4$}
\put(50,110){\circle*{3}}
\put(70,110){\circle*{3}}
\put(90,110){\circle*{3}}
\put(110,110){\circle*{3}}
\put(130,110){\circle*{3}}
\put(2,130){$E_0+5$}
\put(50,130){\circle*{3}}
\put(70,130){\circle*{3}}
\put(90,130){\circle*{3}}
\put(110,130){\circle*{3}}
\put(130,130){\circle*{3}}
\put(150,130){\circle*{3}}
%
%%%%%%%
%%%%%%%%%%%%%%%%%%%%%%%%%%%%%%%%%%%%%%%
%
\end{picture}
\caption{States of the massless $s=1$ representation in terms of the energy
  eigenvalues $E$ and the angular momentum $j$. Now $E_0$ is no longer
arbitrary but it is fixed to $E_0=2$. 
} 
%\vspace{-4mm}
\end{figure}
%%%%%%%%%%%%%%%%%%%%%%%%%%%%%%%%%%%%%%%%%%%%%%%%%%%%%%%%%%%%%

One can try and use the same argument again to see if there is a
possibility that even more states decouple. Consider for instance
a state with the same spin as the ground state, with energy $E$. In
that case 
\be
E(E-3) = E_0(E_0-3) + \Big\vert M_a^- \vert E,s\rangle
\Big\vert^2\,. \label{equal-s-bound} 
\ee
For spin $s\geq 1$, this condition is always satisfied in view of 
the bound $E_0\geq s+1$. But for $s=0$, one can apply
\eqn{equal-s-bound} for
the first excited $s=0$ state which has $E=E_0+2$. In that case one
derives   
\be
2(2E_0 -1) = \Big\vert M_a^- \vert E_0+2,s=0\rangle \Big\vert^2\,,
\ee
so that 
\be
E_0\geq \ft12\,. 
\ee
For $E_0=\ft12$ we have the so-called singleton representation, where
we have only one state for a given value of the spin. A similar result
can be 
derived for $s=\ft12$, where one can consider the first excited state
with $s=\ft12$, which has $E=E_0 +1$. One then derives 
\be
2(E_0 -1) = \Big\vert M_a^- \vert E_0+1,s=\ft12 \rangle \Big\vert^2
\,, 
\ee
so that 
\be
E_0\geq 1 \,.
\ee
For $E_0=1$ we have the spin-$\ft12$ singleton representation, where
again we are left with just one state for every spin value. The
existence of these singleton representations was first noted by
Dirac \cite{Dirac2}. They are shown in Fig.~5. 
Both singletons have the same value of the Casimir operator, 
\be
{\cal C}_2 = -\ft54\,.
\ee
%%%%%%%%%%%%%%%%%%%%%%%%%%%%%%%%%%%%%%%%%%%%%%%%%%%%%%%%%%
%%%%%%%%%%%%%%%%%%%%%% singletons %%%%%%%%%%%%%%%%%%%%%%%%
%%%%%%%%%%%%%%%%%%%%%%%%%%%%%%%%%%%%%%%%%%%%%%%%%%%%%%%%%%
\setlength{\unitlength}{0.5mm}
\begin{figure}[t]
\begin{picture}(260,160)(0,0)
\put(30,10){\vector(1,0){140}}
\put(175,8){$j$}
\put(30,10){\vector(0,1){140}}
\put(27,154){$E$}
\put(28,1){$0$}
\put(48,1){$1$}
\put(68,1){$2$}
\put(88,1){$3$}
\put(108,1){$4$}
\put(128,1){$5$}
\put(148,1){$6$}
\put(50,10){\line(0,1){4}}
\put(70,10){\line(0,1){4}}
\put(90,10){\line(0,1){4}}
\put(110,10){\line(0,1){4}}
\put(130,10){\line(0,1){4}}
\put(150,10){\line(0,1){4}}
\put(30,30){\line(1,0){4}}
\put(30,50){\line(1,0){4}}
\put(30,70){\line(1,0){4}}
\put(30,90){\line(1,0){4}}
\put(30,110){\line(1,0){4}}
\put(30,130){\line(1,0){4}}
\put(2,30){$E_0=\ft12$}
\put(30,30){\circle*{3}}
\put(2,50){$E_0+1$}
\put(50,50){\circle*{3}}
\put(40,40){\circle{3}}
\put(2,70){$E_0+2$}
\put(70,70){\circle*{3}}
\put(60,60){\circle{3}}
\put(2,90){$E_0+3$}
\put(90,90){\circle*{3}}
\put(80,80){\circle{3}}
\put(2,110){$E_0+4$}
\put(110,110){\circle*{3}}
\put(100,100){\circle{3}}
\put(2,130){$E_0+5$}
\put(130,130){\circle*{3}}
\put(120,120){\circle{3}}
%
%%%%%%%%%%%%%%%%%%%%%%%%%%%%%%%%%%%%%%%%%%%%%%
%%%%%%%%%%%%%%%%%%%%%%%%%%%%%%%%%%%%%%%%%%%%%%
%
\end{picture}
\caption{The spin-0 and spin-$\ft12$ singleton representations. The
solid dots indicate the states of the $s=0$ singleton, the circles the
states of the $s=\ft12$ singleton. It is obvious that singletons
contain much less degrees of freedom than a generic local field. The
value of $E_0$, which denotes the spin-0  
ground state energy, is equal to $E_0=\ft12$. The $s=\ft12$ singleton 
ground state has an energy equal to unity, as is explained in the
text. 
} 
%\vspace{-4mm}
\end{figure}
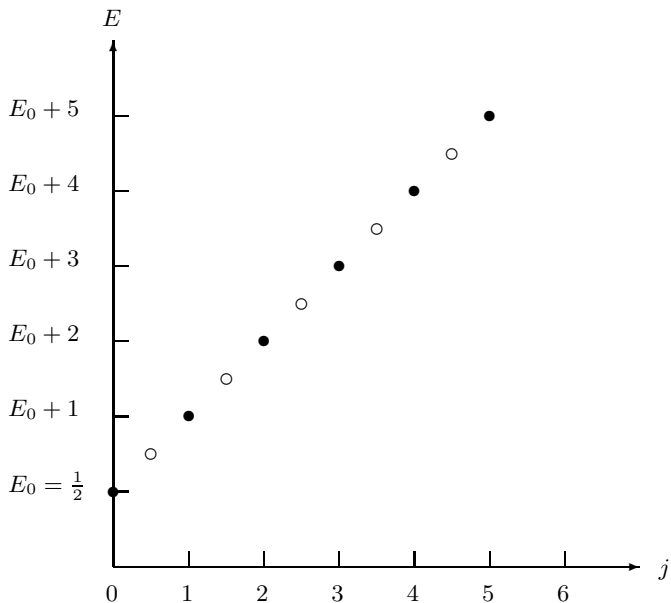
%%%%%%%%%%%%%%%%%%%%%%%%%%%%%%%%%%%%%%%%%%%%%%%%%%%%%%%%%%%%%%

{From} the above it is clear that we are dealing with the phenomenon 
of multiplet 
shortening for specific values of the energy and spin of the
ground state. This can be understood more generally from the fact that
the $[M_a^+,M_b^-]$ commutator acquires zero or negative eigenvalues
for certain values of $E_0$ and $s$. We will return to this phenomenon
in section~\ref{superalgebra} in the context of the anti-de Sitter
superalgebra. 

%%%%%%%%%%%%%%%%%%%%%%%%%%%%%%%%%%%%%%%%%%%%%%%%%%%%%%%%
\section{The oscillator construction}
There exists a constructive procedure for determining the unitary
irreducible representations of the anti-de Sitter algebra, which is
known as the oscillator method. This method can be used
for any number of dimensions and also for the supersymmetric extension
of the anti-de Sitter algebra \cite{GunaSacl,Gunaydin}. 
Here we will demonstrate it for the case of four spacetime dimensions. 

Consider an even $n=2p$ or an odd $n=2p+1$ number
of bosonic harmonic oscillators, whose creation and annihilation
operators transform as doublets under the compact subgroup U$(1)\times
{\rm SU}(2)$  of the covering group Sp$(4)\cong {\rm SO}(3,2)$.
We introduce pairs of mutually commuting  annihilation operators
$a_i(r)$ and $b_i(r)$ labeled 
by $r=1,\ldots, p$ and an optional annihilation operator $c_i$ when we
wish to consider an odd number of oscillators. The indices $i$ are the
doublet indices associated with 
SU(2). The nonvanishing commutation relations are
\bea
{[}a_i(r) ,a^j(s) ] &=& \d_i{}^j\, \d_{rs}\,,    \nn\\ 
{[}b_i(r) ,b^j(s) ] &=& \d_i{}^j\, \d_{rs} \,,\nn\\
{[}c_i ,c^j ] &=& \d_i{}^j\, ,
\eea
where the creation operators carry upper SU(2) indices and are defined
by $a^i= (a_i)^\dagger$, $b^i= (b_i)^\dagger$ and $c^i= (c_i)^\dagger$.
The generators of U$(1)\times
{\rm SU}(2)$ are then given by 
\be
U^i{}_j = a^i\cdot a_j + b_j\cdot b^i + \ft12 (c^i\,c_j + c_j\,c^i)\,, 
\ee
where $a_i\cdot a^j$ stands for $\sum_r a_i(r)\,a^j(r)$. 
The U(1) generator will be denoted by $Q=\ft12 U^i{}_i$ and can be
expressed as  
\bea
Q&=& \ft12(a^i\cdot a_i + b_i\cdot b^i + \ft12 c^ic_i + \ft
12 c_ic^i)\nn\\
&=& \ft12 (a^i\cdot a_i + b^i\cdot b_i + c^ic_i + 2p+1)\nn\\
&=& \ft 12 (N +  n)\,, 
\eea
where $N$ is the number operator for the oscillator states. 
Observe that $Q$ is associated with the generator that we previously
identified with the energy operator. 
The other generators, transforming according to the ${\bf 3} + \bar
{\bf 3}$ representation of SU(2), are defined by 
\be
S^{ij} = (S_{ij})^\dagger = a^i\cdot b^j + a^j\cdot b^i +c^i\,c^j  \,.
\ee
It is now easy to identify the raising and lowering operators by
considering the commutation relations of $Q$ with all the other
operators, 
\be 
{[}Q\,,U^i{}_j] = 0\,,\qquad 
{[}Q\,, S^{ij}] = S^{ij} ,\qquad {[}Q\,, S_{ij}] =
-S_{ij} \,.
\ee
Together with 
\bea
{[} S^{ij},S^{kl}]&=& {[} S_{ij},S_{kl}]= 0\,, \nn\\
{[}S^{ij},S_{kl}] &=& \d^i{}_k \,U^j{}_l +\d^i{}_l \,U^j{}_k +\d^j{}_k
\,U^i{}_l +\d^j{}_l \,U^i{}_k \,,
\eea
we recover all commutation relations of SO($3,2$). 
Obviously, the operators $S^{ij}$ raise the eigenvalue of $Q$, when
acting on its eigenstates, while their hermitian conjugates $S_{ij}$
lower the eigenvalue. Let us, for the sake 
of completeness, write down the commutation relations of $Q$ with the
oscillators,
\be
{[}Q\,, a^i{]}  =  \ft12 a^i \,, \qquad 
{[}Q\,, a_i{]}  =  -\ft12 a_i \,.
\ee
We see that $a^i$ raises the energy by half a unit whereas $a_i$ lowers it 
by the same amount.
The same relations hold of course true for the oscillators $b^i$ and $c^i$.
The ground state $|\O\rangle$ is then defined by 
\be
S_{ij} \vert \Omega\rangle = 0\,. 
\ee
The representation is built by acting with an arbitrary product of raising
operators $S^{ij}$ on the ground state.
Depending on the number of oscillators we have chosen certain states
will be present whereas others will not. 
In this way the shortening of the multiplets
will be achieved automatically. Experience has taught us that the
oscillator construction is complete in the sense that it yields all
unitary irreducible representations. However, it is not possible to
describe the construction for arbitrary dimension, as every case has
its own characteristic properties. 

The obvious choice for $\vert \Omega\rangle$ is the vacuum
state $\vert 0\rangle$ of the oscillator algebra. However, there are other
possibilities. For example, we can act on $\vert0\rangle$
by any number of different creation operators, i.e. 
$a^i(r_1)\,a^j(r_2)\,b^k(r_3)\cdots \vert0\rangle$, as long as we do
not include a pair $a^i(r_1) \, b^j(r_2)$ with $r_1=r_2$, unless it is
anti-ssymmetrized in indices $i$ and $j$. The reason is that $S_{ij}$
consists of terms that 
are linear in both $a_i(r_1)$ and $b_j(r_2)$ annihilation
operators with $r_1=r_2$ and with symmetrized SU(2) indices.
Let us now turn to a number of relevant examples in order to clarify the
procedure.

Assume that we have 
a single harmonic oscillator (i.e. $n=1$). Then there are two
possible ground states. One is $|\Omega\rangle = \vert 0\rangle$. In that
case we have $E_0=Q=\ft12$ and $s=0$. The states take the form of
products of {\it even} numbers of creation operators,
i.e. $c^i\,c^j\,c^k\cdots\vert 
0\rangle$, which are symmetric in the SU(2) indices because the
creation operators are mutually commuting. Obviously these states
comprise states of spin 1, 2, 3, \ldots with multiplicity one.
This is the $s=0$ singleton representation. 
The spin-$\ft12$ singleton follows from choosing the ground state
$|\O\rangle = c^i\vert 0\rangle$, which has $E_0= Q = 1$ and
$s=\ft12$. The states are again generated by even product of creation
operators which lead to states of  spin $\ft32$, $\ft52$, \ldots with
multiplicity one. 

Let us now consider the case of two oscillators ($n=2$). Here we
distinguish the following ground states and corresponding irreducible
representations:
\begin{itemize}
\item One obvious ground state is the oscillator ground state, $\vert
\O\rangle = \vert0\rangle$. In that case we have
$E_0=Q=1$ and $s=0$. This is the massless $s=0$ representation. 
\item Alternative
ground states are $\vert\O\rangle= a^i\vert0\rangle$ or $\vert\O\rangle=
b^i\vert0\rangle$. In that case the ground state has $E_0= Q= \ft32$
and $s=\ft12$. This is the massless $s=\ft12$ representation. 
\item Yet another option is to choose $\vert\O\rangle$ equal
to $m$ annihilation operators exclusively of the $a$-type or of the
$b$-type, applied to $\vert0\rangle$. This ground state has $E_0=Q=1+
\ft12m$ and $s = \ft12 m$. From the values of $E_0$ and $s$ one
deduces that these are precisely the massless spin-$s$
representations. 
\item Finally one may  choose $\vert\O\rangle =
(a^i\,b^j-a^j\,b^i)\vert 0\rangle$, which has $E_0=Q=2$ and
$s=0$. This is the second massless $s=0$ representation.
\end{itemize}

To sum up, for a single oscillator one recovers the singleton
representations 
and for two oscillators one obtains all massless representations. The
excited states in a given representation  are constructed by applying
arbitrary products of an 
{\it even} number of creation operators on the ground state. For more
than two oscillators, one obtains the massive representations. This
pattern, sometimes with small variations, repeats itself for other
than four spacetime dimensions. 

%%%%%%%%%%%%%%%%%%%%%%%%%%%%%%%%%%%%%%%%%%%%%
\section{The superalgebra ${\rm OSp}(1\vert 4)$} 
\label{superalgebra} 
In this section we return to the anti-de Sitter superalgebra. 
We start from the (anti-)commutation relations
already established in \eqn{ads-dec-algebra} and
\eqn{ads-dec-superalgebra}. For definiteness we discuss the case of
four spacetime dimensions with a Majorana supercharge $Q$. This
allows us to make contact with the material discussed in
section~\ref{masslike}. These anti-de Sitter multiplets were discussed
in \cite{BreitFreed,Heidenreich,FreedNicolai,Nicolai}. 

We choose conventions where the gamma matrices are given by
\be
\G^0= \pmatrix{-i {\bf 1}& 0 \cr \noalign{\vskip 2mm}
                0& i {\bf 1} \cr} \,,\qquad 
\G^a= \pmatrix{0 & -i\sigma^a \cr \noalign{\vskip 2mm}
                i\sigma^a & 0 \cr} \,, \quad a=1,2,3 \,,
\ee
and write the Majorana spinor $Q$ in the form
\be
Q= \pmatrix{q_\a \cr \noalign{\vskip 2mm} \varepsilon_{\a\b} \,q^\b
\cr}\,, 
\ee
where $q^\a \equiv q_\a^\dagger$ and the indices $\a,\b,\ldots$ are
two-component spinor indices.
We substitute these definitions into \eqn{ads-dec-superalgebra} and obtain
\bea
{[}H\,,q_\a]  &=& -\ft12 q_\a    \,,\nn\\[1mm]
{[}H\,,q^\a]  &=& \ft12 q^\a    \,,\nn\\[1mm]
\{q_\a\,,q^\b \}  &=& (H\,{\bf 1}+ \vec J\cdot \vec\sigma)_\a{}^\b
\,,\nn\\[1mm] 
\{q_\a\,,q_\b \}  &=& M^-_a \,(\s^a\s^2)_{\a\b}  \,, \nn\\[1mm]
\{q^\a\,,q^\b \}  &=& M^+_a \,(\s^2\s^a)^{\a\b}   \,,
\label{q-sup-alg}
\eea
where we have defined the angular momentum operator 
$J_a = -\ft12\, i\, \varepsilon_{abc} M^{bc}$. We see that the operators
$q_\a$ and $q^\a$ are lowering and raising operators, respectively. They
change the energy of a state by half a unit.

In analogy to the bosonic case, we study unitary irreducible representations 
of the ${\rm OSp}(1\vert 4)$ superalgebra. We assume that there exists
a lowest-weight state $|E_0,s\rangle$, characterized by the fact that it is
annihilated by the lowering operators $q_\a$,
\be
q_\a |E_0,s\rangle = 0 \,.
\ee
In principle we can now choose a ground state and build the whole
representation upon it by applying products of raising operators $q^\a$.
However, we only have to study the {\it antisymmetrized} products of
the $q^\a$, because the symmetric ones just yield products of the
operators $M^+_a$ by virtue of (\ref{q-sup-alg}). Products of the
$M^+_a$ simply lead to the higher-energy states in the anti-de
Sitter representations 
of given spin that we considered in section~\ref{adSreps}. By
restricting ourselves to the antisymmetrized products of the $q^\a$ we 
thus restrict ourselves to   
the ground states upon which the full anti-de Sitter
representations are build. These ground states are   
$|E_0,s\rangle$, $q^\a |E_0,s\rangle$ and
$q^{[\a}q^{\b]}|E_0,s\rangle$. 
Let us briefly discuss these representations for different $s$.

The $s=0$ case is special since it contains less anti-de Sitter
representations than the
generic case. It includes the spinless states $|E_0,0\rangle$ and
$q^{[\a}q^{\b]}|E_0,0\rangle$ with ground-state energies $E_0$ and $E_0+1$,
respectively. There is one spin-$\ft12$ pair of ground states
$q^\a|E_0,0\rangle$, with energy 
$E_0+\ft12$. As we will see below, these states correspond exactly to
the scalar field 
$A$, the pseudo-scalar field $B$ and the spinor field $\psi$ of the chiral 
supermultiplet, that we studied in section \ref{masslike}.

For $s\geq\ft12$ we are in the generic situation. We obtain the ground
states 
$|E_0,s\rangle$ and $q^{[\a}q^{\b]}|E_0,s\rangle$
which have both spin $s$ and which have energies $E_0$ and $E_0+1$,
respectively. There are two more (degenerate) ground states,
$q^\a|E_0,s\rangle$, 
both with energy $E_0+\ft12$, which decompose into the ground states
with spin $j=s-\ft12$ and $j=s+\ft12$.

As in the purely bosonic case of section \ref{adSreps}, 
there can be situations in which states decouple so that we are
dealing with multiplet shortening associated with gauge
invariance in the corresponding field theory. The corresponding 
multiplets are then again called massless. We now discuss this in a
general way analogous to the way in which one discusses BPS multiplets
in flat space. Namely, we consider the 
matrix elements of the operator $q_\a\,q^\b$ between the
$(2s+1)$-degenerate ground states $|E_0,s\rangle$,  
\bea
\langle E_0,s| \,q_\a q^\b\,|E_0,s\rangle & = & 
\langle E_0,s| \{q_\a\,, q^\b\}|E_0,s\rangle \nn \\
& = & \langle E_0,s| (E_0\, {\bf 1}+\vec J \cdot \vec \sigma)_\a{}^\b
|E_0,s\rangle \,. \label{matrix} 
\eea
This expression constitutes an hermitean  matrix in both the
quantum numbers of the 
degenerate groundstate and in the indices $\a$ and $\b$, so that it
is $(4s+2)$-by-$(4s+2)$. Because we 
assume that the representation is unitary, this matrix must be
positive definite, as one can verify by inserting a complete set of
intermediate states between the operators $q_\a$ and $q^\b$ in the
matrix element on the left-hand side. 
Obviously, the right-hand side is manifestly hermitean as well, but in
order to be positive definite the eigenvalue $E_0$ of $H$ must be
big enough to compensate for possible negative eigenvalues of $\vec
J\cdot\vec\sigma$, where the latter is again  
regarded as a $(4s+2)$-by-$(4s+2)$ matrix. To determine its 
eigenvalues, we note that 
$\vec J\cdot\vec\sigma$ satisfies the following identity, 
\be
(\vec J\cdot \vec\sigma)^2 + (\vec J\cdot \vec\sigma) = s(s+1) {\bf 1} \,,
\ee
as follows by straightforward calculation.  
This shows that $\vec J\cdot\vec\sigma$ has only two (degenerate)
eigenvalues (assuming $s\not =0$, so that the above equation is not
trivially satisfied), namely $s$ and $-(s+1)$. Hence in
order for \eqn{matrix} to be positive definite, $E_0$ must satisfy the
inequality 
\be
E_0 \geq s+1 \,, \quad \mbox{for } s \geq \ft12 \,,
\ee
If the bound is saturated, i.e.\ if $E_0 = s+1$, the 
expression on the right-hand side of \eqn{matrix} has zero eigenvalues
so that there are zero-norm states in the multiplet which
decouple. In that case we must be dealing with a massless
multiplet. As an example we mention the case $s=\ft12, 
E_0=\ft32$, which corresponds to the massless vector supermultiplet in
four spacetime dimensions. Observe that we have multiplet shortening
here without the presence of central charges. 

One can also use the oscillator method discussed in the previous section to
construct the irreducible representations. This is, for instance, done
in \cite{GunaWarn,GunaNieuWarn}.

Armed with these results we return to the masslike terms of section
\ref{masslike} for the chiral supermultiplet. The ground-state energy
for anti-de Sitter multiplets corresponding to the scalar field $A$, the
pseudo-scalar field $B$ and the Majorana spinor field  $\psi$, are
equal to $E_0$, $E_0+1$ and
$E_0+\ft12$, respectively. The Casimir operator therefore takes the values
\bea
{\cal C}_2{(A)} & = & E_0(E_0-3) \,,\nonumber \\
{\cal C}_2{(B)} & = & (E_0+1)(E_0-2)\,,\nonumber \\
{\cal C}_2{(\psi)} & = & (E_0+\ft12)(E_0-\ft52) +\ft34 
\label{chiralcasimir}\,.
\eea
For massless anti-de Sitter multiplets, we know that the quadratic
Casimir operator is given by \eqn{masslesscasimir}, so we present the
value for ${\cal C}_2 - 2(s^2-1)$ for the three multiplets, i.e
\bea
{\cal C}_2{(A)} +2 & = & (E_0-1)(E_0-2)\,,\nonumber\\
{\cal C}_2{(B)} +2& = & E_0(E_0-1)\,,\nonumber\\
{\cal C}_2{(\psi)} +\ft32 & = & (E_0-1)^2 \,.
\eea
The terms on the right-hand side are not present for massless fields
and we should therefore identify them somehow with the common mass
parameter. Comparison with the field equations \eqn{field-eqs} shows
for $g=1$ 
that we obtain the correct contributions provided we make the
identification $E_0=m+1$. Observe that we could have made a slightly
different identification here; the above result remains the same under
the interchange of $A$ and $B$ combined with a change of sign in $m$
(the latter is accompanied by a chiral redefinition of $\psi$). 

Outside the context of supersymmetry, we could simply assign
independent mass terms with a mass parameter $\mu$ for each of the
fields, by equating ${\cal C}_2 
- 2(s^2-1)$ to $\mu^2$. In this way we obtain
\be
E_0(E_0-3) - (s+1) (s-2) = \mu^2\,,
\ee
which leads to 
\be
E_0 = \ft 32 \pm  \sqrt{ (s-\ft12)^2 + \mu^2} \,. \label{E0-eq}
\ee
For $s\geq \ft12$ we must choose the plus sign in \eqn{E0-eq} in order to
satisfy the unitarity bound $E_0\geq s+1$. For $s=0$ 
both signs are acceptable as long as $\mu^2\leq \ft34$. 
Observe, however,  that $\mu^2$ can be negative but remains
subject to the 
condition $\mu^2\geq - (s-\ft12)^2$ in order that $E_0$ remains
real. For $s=0$, this is precisely the bound of Breitenlohner
and Freedman for the stability of the anti-de Sitter background
against small fluctuations of the scalar fields \cite{BreitFreed}.  

We can also compare ${\cal C}_2 - 2(s^2-1)$ to the
conformal wave operator for the corresponding spin. This shows that 
(again with unit anti-de Sitter radius), ${\cal C}_2 = \Box_{\rm adS} +
\d_{\rm s}$, where $\d_{\rm s}$ is a real number depending on the spin
of the field. Comparison with the field equations of
section~\ref{masslike} shows that $\d_{\rm s}$ equals 0, $\ft32$ and
3, for $s=0,\ft12$ and 1, respectively. 

In the case of $N$-extended supersymmetry the supercharges transform
under an SO($N$) group and we are dealing with the so-called OSp($N|4$)
algebras. Their representations can be constructed by the methods
discussed in these lectures. However, the generators of SO($N$) will now
also appear on the right-hand side of the anticommutator of the two
supercharges, thus leading to new possibilities for multiplet
shortening. For an explicit discussion of this we refer
the reader to \cite{FreedNicolai}.

\section{Conclusions} 
In these lectures we discussed the irreducible representations of the
anti-de Sitter algebra and its superextension. Most of our discussion
was restricted to four spacetime dimensions, but in principle the same
methods can be used for anti-de Sitter spacetimes of arbitrary
dimension. 

For higher-extended supergravity, the only way to
generate a cosmological constant is by elevating a subgroup of the rigid
invariances that act on the gravitini to a local group. This then leads to a
cosmological constant, or to a potential with possibly a variety of
extrema, and corresponding masslike terms which are quadratic and
linear in the gauge coupling constant, respectively. So the relative
strength of the anti-de 
Sitter and the gauge group generators on the right-hand side of the
$\{Q,\bar Q\}$ anticommutator is not arbitrary and because of that
maximal multiplet shortening can take place so that the theory can
realize 
a supermultiplet of massless states that contains the graviton and the
gravitini. Of course, this is all under the assumption that the ground
state is supersymmetric. But these topics are outside the scope of
these lectures and will be reviewed elsewhere \cite{trieste}. 
 
\vspace{4mm}

\noindent
We thank M. G\"unaydin for valuable comments. 
IH is supported by the Swiss National Science Foundation through the
graduate  fellowship 83EU-053229. This work is also supported by the
European Commission TMR programme ERBFMRX-CT96-0045.

%%%%%%%%%%%%%%%%%%%%%%%%%%%%%%%%%%%%%%%%%%%%%%%%%%%%%%%%%%%%%%%%%%%%%%%
%%%%%%%%%%%%%%%%%%%%%%%%%%%%%%%%%%%%%%%%%%%%%%%%%%%%%%%%%%%%%%%%%%%%%%%%
%%%%%%%%%%%%%%%%%%%%%%%%%%%%%%%%%%%%%%%%%%%%%%%%%%%%%%%%%%%%%%%%%%%%%%%%%

%INDEX%%%%%%%%%%%%%%%%%%%%%%%%%%%%%%%%%%%%%%%%%%%%%%%%%%%%%%%%%%%%%%%
%\clearpage
%\addcontentsline{toc}{section}{Index}
%\flushbottom
%\printindex
%%%%%%%%%%%%%%%%%%%%%%%%%%%%%%%%%%%%%%%%%%%%%%%%%%%%%%%%%%%%%%%%%%%%%


\begin{thebibliography}{30}
%
\addcontentsline{toc}{section}{References}
%
\bibitem{Mald} 
  J. Maldacena, The large-$N$ 
  limit of superconformal field theories and supergravity,
  Adv. Theor. Math. Phys. {\bf 2} (1998) 231, {\tt hep-th/9711200}. 
%
\bibitem{Dirac1} P.A.M. Dirac, The electron wave equation in de Sitter space,
Ann. Math. {\bf 36} (1935) 657.
%
\bibitem{Dirac2} P.A.M. Dirac, 
  A remarkable representation of the $3+2$
  de Sitter group,  J. Math. Phys. {\bf 4} (1963) 901.
%
\bibitem{Fronsdal} C. Fronsdal, Elementary particles in a curved space, 
  Rev. Mod. Phys. {\bf 37} (1965) 221;
  Elementary particles in a curved space II, 
  Phys. Rev. {\bf D10} (1974)   589; C. Fronsdal and R.B. Haugen, 
  Elementary particles in a curved
  space III, Phys. Rev. {\bf D12} (1975) 3810; 
  C. Fronsdal, Elementary particles in a
  curved space IV, Phys. Rev. {\bf D12} (1975) 3819.  
%
\bibitem{AvisIshamStorey} S.J. Avis, C.J. Isham and D. Storey, Quantum
  field theory in anti-de Sitter space-time, Phys. Rev. {\bf D18} (1978)
3565. 
%
\bibitem{Dusedau:1986ue} D.W.~D\"usedau and D.Z.~Freedman,
  Lehmann spectral representation for anti-de Sitter quantum field theory,
  Phys. Rev. {\bf D33} (1986) 389.
  %%CITATION = PHRVA,D33,389;%%
%
\bibitem{DZF} D.Z.~Freedman, Supergravity with axial-gauge invariance,
Phys. Rev. {\bf D15} (1977) 1173.
%
\bibitem{FDas} 
  D.Z.~Freedman and A.~Das, Gauge internal symmetry in extended supergravity,
  Nucl. Phys. {\bf B120} (1977) 221.
%
\bibitem{FSchwarz}
  D.Z.~Freedman and J.H.~Schwarz, N=4 supergravity theory with local SU(2)
  $\times$ SU(2) invariance, Nucl. Phys. {\bf B137} (1978) 333.
%
\bibitem{dWNic} B. de Wit and H. Nicolai, Extended supergravity
with local SO(5) invariance, Nucl. Phys. {\bf B188} (1981) 98.
%
\bibitem{deWitNic}
  B. de Wit and H. Nicolai, $N=8$ supergravity with local SO(8)
  $\times$ SU(8) invariance, Phys. Lett. {\bf 108B} (1982)
  285; $N=8$ supergravity, Nucl. Phys. {\bf B208} (1982) 323.
%
\bibitem{GatesZ} S.J. Gates and B. Zwiebach, Gauged $N=4$ supergravity
with a new scalar potential, Phys. Lett. {\bf 123B} (1983) 200.
%
\bibitem{PPvN}
  M. Pernici, K. Pilch and P. van Nieuwenhuizen, Gauged maximally extended 
  supergravity in seven dimension, Phys. Lett. {\bf 143B} (1984) 103. 
%
\bibitem{Hull} C.M. Hull, More gaugings of $N=8$ supergravity,
Phys. Lett. {\bf 148B} (1984) 297.
%
\bibitem{GianPernN}
  F. Gianni, M. Pernici and P. van Nieuwenhuizen, Gauged $N=4$ supergravity 
  in six dimension, Phys. Rev. {\bf D30} (1984) 1680.
%
\bibitem{GunaRomansWarner}
  M. G\"unaydin, L.J. Romans and N.P. Warner,
  Compact and non-compact gauged supergravity theories in five
  dimensions, Nucl. Phys. {\bf B272} (1986) 598. 
%
\bibitem{BreitFreed} P. Breitenlohner and D.Z. Freedman, Stability in
  gauged extended supergravity, Ann. Phys. {\bf 144} (1982) 249.
%
\bibitem{Heidenreich} 
  W. Heidenreich, All linear unitary irreducible representations of de Sitter
  supersymmetry with positive energy, Phys. Lett. {\bf 110B} (1982) 461.
%
\bibitem{FreedNicolai} D.Z. Freedman and H. Nicolai, Multiplet shortening
  in OSp$(N|4)$,  Nucl. Phys. {\bf B237} (1984) 342.
%
\bibitem{Nicolai} H. Nicolai, Representations of supersymmetry in
  anti-de Sitter space, in Supersymmetry and 
  Supergravity '84, Proceedings of the Trieste Spring School, 
  eds. B. de Wit, P. Fayet, P. van Nieuwenhuizen 
  (World Scientific, 1984).
%
\bibitem{GunaWarn} M. G\"unaydin and N.P. Warner, Unitary
  supermultiplets of OSp$(8|4,{\bf R})$ and the spectrum of the $S^7$
  compactification of eleven-dimensional supergravity, Nucl. Phys. {\bf B272}
  (1986) 99. 
%
\bibitem{GunaNieuWarn} M. G\"unaydin, P. van Nieuwenhuizen and
  N.P. Warner, General construction of the unitary representations of anti-de
  Sitter superalgebras and the spectrum of the S$^4$ compactification of
  eleven-dimensional supergravity, Nucl. Phys. {\bf B255} (1985) 63.
%
\bibitem{Ferrara} S. Ferrara, Algebraic properties of extended
supergravity in de Sitter space, Phys. Lett. {\bf 69B} (1977) 481.
%
\bibitem{deWitZwartk} B. de Wit and A. Zwartkruis, SU($2,2|1,1$) 
  supergravity and $N=2$ supersymmetry with arbitrary cosmological constant, 
  Class. Quantum Grav. {\bf 4} (1987) L59.
%
\bibitem{townsend} P.K. Townsend, Cosmological constant in
  supergravity, Phys. Rev. {\bf D15} (1977) 2802.
%
\bibitem{deser-zumino} S. Deser and B. Zumino, Broken supersymmetry
  and supergravity, Phys. Rev. Lett. {\bf 38} (1977) 1433.
%
\bibitem{nahm} W. Nahm, Supersymmetries and their representations,
Nucl. Phys. {\bf B135} (1978) 149.
%%
\bibitem{deWit82}
  B. de Wit, Multiplet calculus, in Supersymmetry and Supergravity '82,
  Proceedings of the Trieste School, eds. S.~Ferrara, J.G.~Taylor and 
  P.~Van Nieuwenhuizen (World Scientific, 1983).
%
\bibitem{GunaSacl} M. G\"unaydin and C. Saclioglu, Oscillator like unitary
  representations of noncompact groups with a Jordan structure and the 
  noncompact groups of supergravity, Commun. Math. Phys. {\bf 91} (1982) 159.
%
\bibitem{Gunaydin}
  M. G\"unaydin, Oscillator like unitary representations of noncompact groups
  and supergroups and extended supergravity theories, in Int. Colloq. on
Group Theoretical Methods in Physics, Istanbul 1982,
eds. M.~Serdaroglu and E.~In\"on\"u,  
  Lect. Notes in Physics {\bf 180} (Springer, 1983).
%
\bibitem{trieste}
  B. de Wit, Gauged supergravity, lectures at the Trieste Spring School 1999, 
  to be published in the proceedings.
%
\end{thebibliography}
\end{document}